\documentclass[conference]{IEEEtran}
\IEEEoverridecommandlockouts
\usepackage{cite}
\usepackage{makecell}

\usepackage{amsmath,amssymb,amsfonts}
\usepackage{amsmath}
\usepackage{graphicx}
\usepackage{textcomp}
\usepackage{xcolor}
\usepackage{subfigure}
\usepackage{algpseudocode} 
\usepackage{multirow}
\usepackage{tablefootnote}
\usepackage{setspace}
\usepackage{amsmath}
\usepackage{times}
\usepackage{soul}
\usepackage{url}
\usepackage{mathrsfs}
\usepackage{amsmath}
\usepackage{amsthm}
\usepackage{amsfonts}
\usepackage{stfloats}
\usepackage{color}
\usepackage{framed}
\usepackage[linewidth=1pt]{mdframed}
\usepackage{booktabs} %调整表格线与上下内容的间隔
\usepackage{graphics,graphicx,color,multirow,subfigure}
\usepackage{enumerate}
\usepackage[hidelinks]{hyperref}
\usepackage[numbers,sort&compress]{natbib}
\usepackage{booktabs} %调整表格线与上下内容的间隔
\usepackage{caption}
\usepackage{makecell}

\def\BibTeX{{\rm B\kern-.05em{\sc i\kern-.025em b}\kern-.08em
    T\kern-.1667em\lower.7ex\hbox{E}\kern-.125emX}}
\begin{document}

\title{Does Money Laundering  on Ethereum Have Traditional Traits?}
%An empirical analysis of UpbitHack event based on complex network.
\author{\IEEEauthorblockN{Qishuang Fu\IEEEauthorrefmark{1},
		Dan Lin\IEEEauthorrefmark{2}, Yiyue Cao\IEEEauthorrefmark{1}, Jiajing Wu\IEEEauthorrefmark{1}\IEEEauthorrefmark{3} 
	}
	\IEEEauthorblockA{\IEEEauthorrefmark{1}School of Computer Science and Engineering, Sun Yat-sen University, Guangzhou 510006, China\\}
	\IEEEauthorblockA{\IEEEauthorrefmark{2}School of Software Engineering, Sun Yat-sen University, Zhuhai 519082, China\\}
    \IEEEauthorblockA{\IEEEauthorrefmark{3}Correponding Author: wujiajing@mail.sysu.edu.cn}
	}

\maketitle

%-------------------摘要---------------------
% 背景：
% %蓬勃发展的以太坊因为匿名特性，吸引了众多犯罪分子做非法活动并通过洗钱来获得收益。[参考洗钱报告] 

% 不足：
% %在传统洗钱场景中，研究人员揭露了洗钱网络的普遍特性。
% 但是，因为以太坊洗钱是一种新兴的手段。little is known about 以太坊洗钱。

% 引出我们的工作：
% %To fill the gap,  in this paper, we conduct a 洗钱network analysis on a 经典的以太坊安全事件--Upbit Hack 去探究区块链上的洗钱网络是否也有传统的特性。

% 具体来说，
%% 通过爬取交易数据构建了 ML-ETH 网络；
%% 根据洗钱网络传统特性提出了five how questions, 通过利用网络分析去characterize以太坊洗钱网络并回答这些问题。
%% 获得了许多有意思的观察和findings，这为未来对区块链的洗钱检测打下了基础。
\begin{abstract}
%--------------1107version-------------------
As the largest blockchain platform that supports smart contracts, Ethereum has developed with an incredible speed. Yet due to the anonymity of blockchain, the
popularity of Ethereum has fostered the emergence of various illegal activities and money laundering by converting ill-gotten funds to cash. 
In the traditional money laundering scenario, researchers have uncovered the prevalent traits of money laundering. However, since money laundering on Ethereum is an emerging means, little is known about money laundering on Ethereum. To fill the gap, in this paper, we conduct an in-depth study on Ethereum money laundering networks through the lens of a representative security event on \textit{Upbit Exchange} to explore whether money laundering  on Ethereum has traditional traits. Specifically, we construct a money laundering network on Ethereum by crawling the transaction records of \textit{Upbit Hack}. Then, we present five questions based on the traditional traits of money laundering networks. By leveraging network analysis, we characterize the money laundering network on Ethereum and answer these questions. In the end, we summarize the findings of money laundering networks on Ethereum, which lay the groundwork for money laundering detection on Ethereum.

%-------------yy-comment-------------------
%(1)In (the) traditional money laundering scenario
%(2) lense?? 
%qs-comment 已改~
\end{abstract}


\section{Introduction}
%P1:区块链和以太坊的背景： 发展迅速
%1.区块链自2008年诞生以来，区块链凭靠着快速增长的应用in various fields席卷世界。
%2.以太坊作为最大的支持智能合约的平台，越来越繁荣，且吸引了全世界的投资者。
%以太坊上的原生币是ETH，其他币种叫做代币。大量投资者可以在以太坊上无成本的创建账户去交易。
%截止目前，以太坊的市场价值已经到达了154B。
Since the debut of Bitcoin in 2008~\cite{nakamoto2008bitcoin}, blockchain, as the underlying technology of Bitcoin, has taken the world by storm with its mushromming applications in various fields. 
As the largest blockchain platform that supports smart contracts~\cite{Lin2021Evolution}, Ethereum becomes prosperous and has attracted lots of investors.
The native cryptocurrency on Ethereum is called Ether (ETH for short) and other cryptocurrencies on Ethereum are  collectively called tokens~\cite{Wu2021JNCA}. Numerous investors register the account without cost on Etherum to make transactions with ETH and tokens~\cite{lin2022ethereum}. %To date, the market capitalization of Ethereum has reached about 154 billion USD\footnote{\url{https://coinmarketcap.com/currencies/Ethereum/}}.

%P2: 引出区块链上的洗钱
%1.以太坊蓬勃发展的同时吸引不法分子进行非法活动。
%根据报告，从2017年到2021年，不法分子收到的总金额为42.5B USD,这给经济造成了严重的影响。
%2.这些数额庞大的黑钱被cybercriminals移到了服务提供商那，然后提现，这就是洗钱的过程。
%3.由于犯罪分子对金钱的贪欲，洗钱几乎是所有犯罪活动的下游活动
%4.自2017年以来，有33 B usd被犯罪分子清洗。
While Ethereum thrives, it attracts criminals to engage in illegal activities because of a lack of powerful law enforcement on Ethereum~\cite{modeling2020lin}. According to a crypto crime report of Chainalysis~\cite{Chainalysis2022}, the total cryptocurrency value received by illicit accounts from 2017 to 2021 is highly reached 42.5 billion USD.%, which makes a serious impact on economics. 
~The ill-gotten money with an astronomical figure is moved to some service providers, such as mixing services, token swap services, and loan services, and then converted to cash by cybercriminals, which is the process of money laundering (ML for short). Generally, ML process consists of 3 parts~\cite{lokanan2022predicting}, placement, layering, and integration. Due to the avarice of criminals for money, ML is the downstream activity of almost all forms of cryptocurrency-based crimes. 
%33 billion USD worth of cryptocurrency from 2017 to 2021 has been laundered by criminals~\cite{Chainalysis2022}.

%-----------yy-comment----------
%(3) (the) total cryptocurrency value received...
%qs-comment 已改
%P3:以太坊反洗钱研究很少，但传统金融场景反洗钱方案很多，而且大多都是基于洗钱特性提出的方法。
%1.以太坊上犯罪分子嚣张的洗钱行为引来了学术界和工业界的关注，反洗钱技术迫切需要被提出去保护以太坊系统不被犯罪分子污染。
%2.由于缺乏对 以太坊洗钱的了解，以太坊反洗钱的研究都还在初步探索阶段。
%3.相反，传统金融场景的反洗钱研究已经很成熟了,多种多样的反洗钱方法基于洗钱特性被提出。
%4.举例：
%ref:
The arrogant ML activity by criminals on Ethereum has drawn the attention of industry~\cite{Chainalysis2022,slowmist2022,CipherTrace2022,fatf2022}, and anti-money laundering (AML for short) techniques urgently need to be proposed to protect the Ethereum ecosystem from being polluted by criminals.
Owing to the lack of understanding of ML on Ethereum, AML research on Ethereum is still in the preliminary stage of exploration. 
On the contrary, AML research in traditional financial scenarios is very mature~\cite{sun2021cubeflow,li2020flowscope,sun2022monlad,mahootiha2021designing,singh2019anti,jullum2020detecting}, and a variety of AML techniques are proposed based on ML characteristics. For example, Sun et al.~\cite{sun2021cubeflow} combines three typical characteristics of ML to design a multi-attribute metric and apply it to a near-greedy algorithm to detect ML transactions in a bank. Li et al.~\cite{li2020flowscope} utilizes two traits of ML to formulate an anomalousness metric and proposes an algorithm called Flowscope to detect dense ML networks. 

%P4: 传统洗钱特性
%1.对洗钱特性有一个充足的了解 有助于提出反洗钱方案。存在的工作总结了5个传统金融场景中的洗钱特性：

Having an adequate understanding of the characteristics of ML paves the way to come up with AML techniques. We summarize  five traits of ML in traditional financial scenarios from existing works~\cite{sun2021cubeflow,li2020flowscope,sun2022monlad,mahootiha2021designing}:
\begin{mdframed}

\textbf{Trait 1.~\cite{sun2021cubeflow}} Traditional ML accounts are fast-in and fast-out accounts. Most of those accounts transfer money in and out within a short interval. 

\noindent
\textbf{Trait 2.~\cite{li2020flowscope}} Traditional ML accounts prefer to transfer high-volume money since the number of accounts is limited. 

\noindent
\textbf{Trait 3.~\cite{sun2022monlad}} Traditional ML accounts  are zero out middle accounts. Most received money of ML accounts will be transferred out.

\noindent
\textbf{Trait 4.~\cite{sun2021cubeflow}} Traditional ML networks are dense. The accounts on ML networks cooperate to trade with one another frequently many times.

%问题：这里的integration需要解释一下吗？
\noindent
\textbf{Trait 5.~\cite{mahootiha2021designing}} Accounts of integration subset on the traditional ML network have high closeness centrality, since the ill-gotten money will concentrate on the accounts of integration subset.
\end{mdframed}
%1.研究人员不禁好奇以太坊ML网络是否存在这些特性呢？
%2.回答这个问题以后，研究人员会更加了解以太坊ML，

Researchers can not help but wonder if ML networks on Ethereum have
these traditional traits. They will have a better understanding of ML networks on Ethereum after acquiring the answer, which provides theoretical foundations to  design and improve ML detection methods on Ethereum and fills the gap of AML on Ethereum.

%P5:我们的工作
%1. 收集了upbit hack的交易作为MLdataset
%2. 通过比较ML和正常的网络，分析了ML的网络性质
%3.挖掘了以太坊ML和传统ML不同的特性。
In this paper, we provide insights into the ML network on Ethereum through the analysis of network properties. We collect the transactions of \textit{Upbit Hack}\footnote{\url{https://upbit.com/service_center/notice?id=1085}} to construct ML network as the example of ML dataset on Ethereum. Then, we conduct analysis on the ML network properties by comparing the ML network with normal network on Ethereum and dig out the traits of ML network on Ethereum which are different or similar with traditional ML networks.

% The remaining parts of the paper is organized as follows. Section \ref{sec:dataset} introduces the process of collecting the ML dataset and constructing  ML networks. Section \ref{sec:analysis} describes definitions of five network properties and presents the analysis of network properties.
% Conclusion and future work are described in Section \ref{sec:conclusion}. 

\section{Dataset and ML Network}
\label{sec:dataset}
\subsection{Data Collection}
%-------------------数据收集---------------------
%【短句】无公开洗钱数据集，所以自行用工具获取数据：
%%目前没有公开的以太坊洗钱数据集供研究者使用，所以我们利用强大的以太坊浏览器Etherscan和高效的数据爬取工具BlockchainSpider去collect 以太坊洗钱数据集。

%【短句】因为在有大量标签，且案件典型，所以我们收集了这个案件的洗钱数据集。
%%在Etherscan的标签云中有大量Upbit Hack标签的账户。UpbitHack案件是一个典型的洗钱案件。臭名远扬的Upbit Hacker在2019 年 11 月 27日盗取了约342,000 ETH，并利用大量的账户分层转移将资金分层转移，最终存入多个交易所中。So，我们利用Blockchain爬取被Etherscan标记为Upbit Hacker的交易数据作为以太坊洗钱数据集。

%层层递进地爬取数据集：
% 第一步：爬取815个账户的交易数据。
% 第二步：将交易按照交易对象账户的标签分成layering transactions and destination transactions
%如果交易对象是服务商，那么交易就是目标交易；如果交易对象不是，那就是layering 交易
%第三步：继续爬取非服务商账户的交易，知道最后一层交易都是目标交易。

%----------------1108version------------------
Since there is no publicly available  ML dataset on Ethereum, we utilize XBlock\footnote{\url{https://www.xblock.pro/cloud}}, a prominent blockchain explorer, and BlockchainSpider~\cite{wu2022transaction}, an open-source crawler toolkit, to collect ML dataset on Ethereum.
In XBlock's Label Word Cloud, there are 815 accounts labeled as \textit{Upbit Hack}. The \textit{UpbitHack} event is a representative event of ML on Ethereum. Notorious \textit{Upbit Hack} stole approximately 342,000 ETH on November 27, 2019, then %leveraged a large number of accounts to 
transferred the funds in layers to various service providers and eventually withdraw funds from those service providers. So we use BlockchainSpider to crawl transaction records of \textit{Upbit Hack} to build an ML dataset on Ethereum.

%------------yy-comment---------
%(4) an open(-)source crawler..
%(5)倒数第二句话有两个“and”
%qs comment 已改~

The ML dataset is crawled layer by layer. In the first step, we crawled transaction records of 815 \textit{Upbit Hack}. In the second step, we divided those transaction records into two categories, layering transactions and integration transactions, based on the labels of receiving accounts of transactions. If receiving accounts are service providers, such as exchanges and mixing services, the transactions are integration transactions, which are signs of successful ML. The transactions of non-service provider accounts are layering transactions, which indicate ML is still going on. Then, we crawled non-service providers' transactions until transactions of the last layer are integration transactions. %based on the labels of accounts trading with  \textit{Upbit Hack}. If trading accounts are service providers, such as exchanges and mixing services, the transactions are integration transactions, which are signs of successful ML. The transactions of non-service provider accounts are layering transactions, which indicate ML is still going on. Then, we crawled non-service providers' transactions until transactions of the last layer are integration transactions.

%https://upbit.com/service_center/notice?id=1085
%https://www.xblock.pro/cloud/#/

\subsection{Network Construction}
%-----------------网络构建--------------
%1.构建网络+网络的边的含义
%2.由于以太坊支持创造代币,所以以太坊上交易,按照币种分类,由ETH交易和代币交易组成

%----------------1108version------------------

Consider a ML network on Ethereum~($\small \textsf{LaunderNet}$  for short) $G$ $
=(N, E)$, where $N$, $E$ represent the set of ML accounts and ML transactions, respectively. Each transaction $e(u,v,w,t,c) \in E$ denotes that at timestamp $t$, account $u$ transfers cryptocurrency $c$ of amount $w$ to account $v$, where $u,v \in N$.   

Because Ethereum allows the creation of tokens, transactions on Ethereum consist of ETH transactions and token transactions categorized by the type of cryptocurrency. To make network analysis more convenient, we divide $\small\textsf{LaunderNet}$ into $\small\textsf{LaunderEtherNet}$ and $\small\textsf{LaunderTokenNet}$. The transactions on  $\small\textsf{LaunderEtherNet}$ only trade in ETH while the transactions on $\small\textsf{LaunderTokenNet}$ only trade in tokens. Table~\ref{tab:Statistics of ML networks} shows the statistics of three networks. We can find the sum of the number of accounts on $\small\textsf{LaunderEtherNet}$ and $\small\textsf{LaunderTokenNet}$ is greater than the number of accounts on $\small\textsf{LaunderNet}$. That is because some accounts both trade in ETH and Tokens.

%介绍用来当做参照去评判以太坊ML网络的特的数据集。
In order to find traits of the ML network on Ethereum, we prepare a normal transaction network~\cite{Lee2020Measurements} and normal accounts~\cite{trans2vec2020Wu} with detailed transactions on Ethereum, which are regarded as  references to judge the traits of the ML network on Ethereum.

\begin{table}[htbp]
\footnotesize
  \centering
  \caption{Statistics of ML networks}
    \begin{tabular}{lrr}
    \toprule
        \multicolumn{1}{l}{\textbf{Networks}} 
          & \multicolumn{1}{l}{\#Accounts} & \multicolumn{1}{l}{\#Transactions} \\
    \midrule
    $\footnotesize\textsf{LaunderEtherNet}$ & 377,912 & 1,627,861 \\
    $\footnotesize\textsf{LaunderTokenNet}$ &   239,736 & 720,319  \\
    $\footnotesize\textsf{LaunderNet}$ &    577,313   & 2,348,180 \\
    \bottomrule
    \end{tabular}%
  \label{tab:Statistics of ML networks}%
\end{table}%

\section{The Analysis of Network Properties}
\label{sec:analysis}
%总说段：
%在本部分，我们通过计算并比较以太坊洗钱网络和普通网络的网络性质，去分析以太坊以前网络是否拥有传统洗钱特性。
In this section, we calculate and compare the network properties of the ML network and normal network on Ethereum to analyze whether the ML network on Etherum has traditional ML traits.
%----------------交易时间间隔-------------------
\subsection{How fast is ETH
transferred in and out on \footnotesize{\textsf{LaunderEtherNet}}?}

%问题：这个caption会有歧义嘛
\begin{figure}[t]
  \centering
  \includegraphics[width=0.8\linewidth]{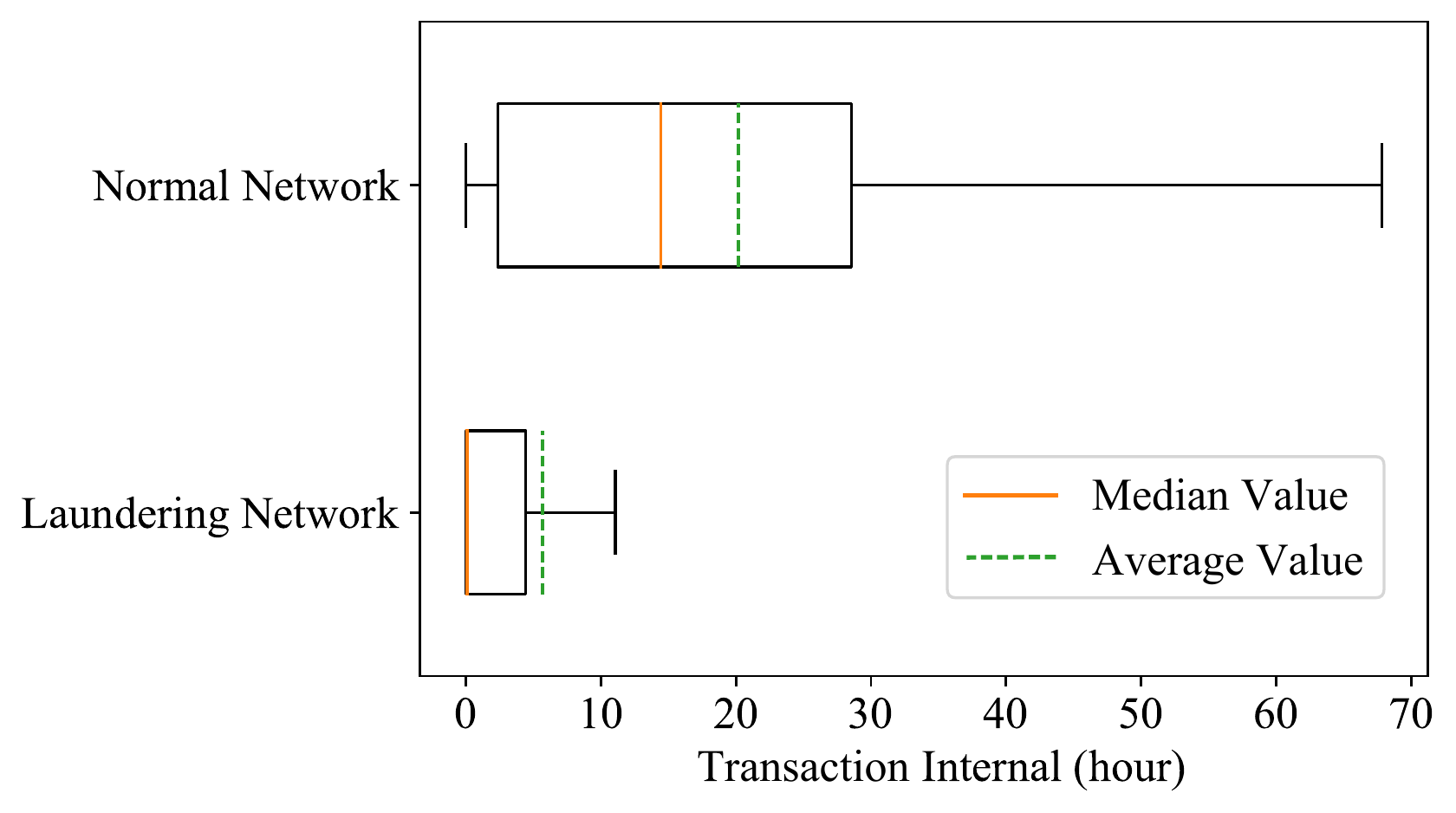}
  \caption{Average transaction interval of account on normal network and money laundering network. }
  \label{fig:Tx Internal}
\end{figure}
%---------------介绍交易间隔指标公式-------------
%为了和Trait1对比去探究以太坊洗钱账户是否是快进快出账户，我们引入了平均交易间隔去measure洗钱账户的交易速度。
%账户的平均交易间隔是+公式，

In order to compare with Trait 1 to explore whether ML accounts on Ethereum are fast-in and fast-out accounts, we introduce the average transaction interval to measure the transaction velocity of the ML accounts. The average transaction interval (Avg. Tx Interval for short) of an account $u$ is defined as 
\begin{equation}
\nonumber
    I(u) = \frac{\sum T(e_{pair}(u))}{N(e_{pair}(u))},(e_{pair} \in E),
\label{equ:reliab}
\end{equation}
where $e_{pair}(u)$ is the transaction pair of $u$. For example, $u$ receives one ETH and then transfers it to other account, and those two transactions are a transaction pair. $T(e_{pair}(u))$ is the interval within a transaction pair of $u$ and $N(e_{pair}(u))$ is the number of 
transaction pairs of $u$.

%-----------------交易间隔结果分析--------------
%Upbit Hacker的交易间隔的均值, 中位数都比Normal Account的小很多，这说明脏钱会被洗钱账户快速转进并快速转出。
%我们推测 这是一种手段去减少监管部门的注意并且更快速地完成洗钱过程得到收益。
Then we calculate Avg. Tx Interval of normal accounts and ML accounts. In Fig.~\ref{fig:Tx Internal}, we present the boxplot of Avg. Tx Interval of normal network and ML network. As shown in Fig.~\ref{fig:Tx Internal}, the mean and median value of Avg. Tx Interval of the ML network are all quite smaller than those of the normal network, which indicates that dirty money is transferred in and out quickly by ML accounts. We can infer that transferring money in and out quickly on Ethereum is a means to reduce the attention of regulators and complete the process of ML  more quickly to gain revenue.

%举交易间隔短的账户例子
%0xEc305a14191fa627B151F16F59183e7E274d81ED upbit hack 7.6
%5个交易对，交易间隔都小于10分钟
\textbf{[Example]} In ML dataset, \textit{Upbit Hack 7.5}\footnote{\scriptsize\url{https://cn.etherscan.com/address/0xec305a14191fa627b151f16f59183e7e274d81ed}} is a typical example with five transaction pairs and the interval with each pair  is less than ten minutes.

% Finding1: 和传统的洗钱账户一样，区块链洗钱网络中的洗钱账户是Fast in and Fast out  Accounts。他们会在短时间内将转入的钱转出，减少被监管的风险。

\begin{mdframed}
\textbf{Finding 1.} Like traditional ML accounts,  ML accounts on Ethereum are fast-in and fast-out accounts. They transfer money in and out in a short period of time to reduce the risk of being regulated.
\end{mdframed}

%------------------交易金额---------------------
\subsection{How high is transfer volume on $\footnotesize\textsf{LaunderEtherNet}$?}
\begin{figure}[t]
  \centering
  \includegraphics[width=0.8\linewidth]{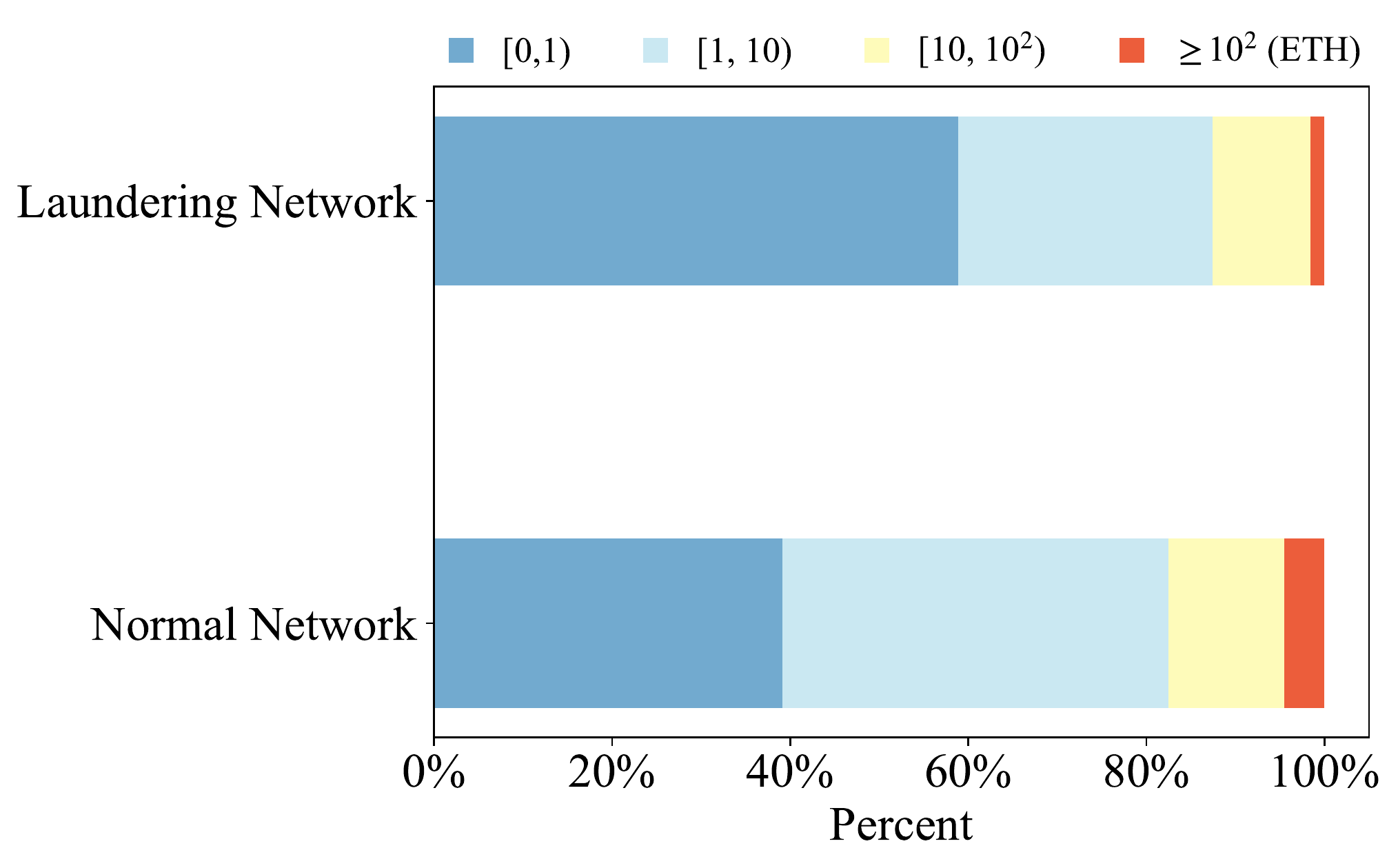}
  \caption{Transaction volume on normal network and money laundering network. }
  \label{fig:Tx value}
\end{figure}

%-----------------交易金额结果分析--------------
%为了分析ML账户的交易金额偏好，我们将交易金额划分为4个范围，然后计算ML账户和普通账户每种范围的占比。
%
%近60%的洗钱交易金额小于1ETH,且金额小于1ETH的占比也大于普通账户的
%不同于传统洗钱账户，以太坊上的洗钱账户不会一次性把大额的虚拟货币转过去，因为这很容易引人注意，洗钱者多次转移小数额的钱。
To analyze the preference of ML accounts for the volume of transactions, We divide the transaction volume into four ranges, then calculate the percentage of each kind range of  the ML network and the normal network. Fig. \ref{fig:Tx value} shows the percentage of transaction volume of ML network and normal network. It can be seen from Fig. \ref{fig:Tx value} that nearly 60\% of the volume of ML transactions are for less than one ETH and the percentage of less than one ETH of ML transactions is larger than that of normal transactions. Unlike traditional 
ML accounts, ML  accounts on Ethereum do not transfer high-volume money  at once, as this will easily attract attention of regulator. Instead, tricky ML accounts transfer very small-volume money multiple times.

\textbf{[Example]} In ML dataset, 0x5217c\footnote{\scriptsize\url{https://cn.etherscan.com/address/0x5217c74fc29fe53c5f75a11cd95f08282fbd97a1}} is a representative example, which received 0.6 ETH and then transferred those ETH with 304 transactions of very small volume, and 85\% of these transactions are for less than 0.001 ETH.

\begin{mdframed}
\textbf{Finding 2.} Compared with traditional ML accounts that usually transfer high-volume money, prudent ML accounts on Ethereum tend to transfer very small-volume money to evade the attention of regulatory authorities. 
\end{mdframed}

%举交易金额小的账户例子
%0x5217c74fc29fe53c5f75a11cd95f08282fbd97a1
%他收到了2笔共0.6ETH，但通过303条极小额交易将这0.6ETH转出
%-----加油------------      
\subsection{How balanced is {\small{ETH}} transferred in and out on \footnotesize\textsf{LaunderEtherNet}?}

\begin{figure}[t]
  \centering
  \includegraphics[width=0.73\linewidth]{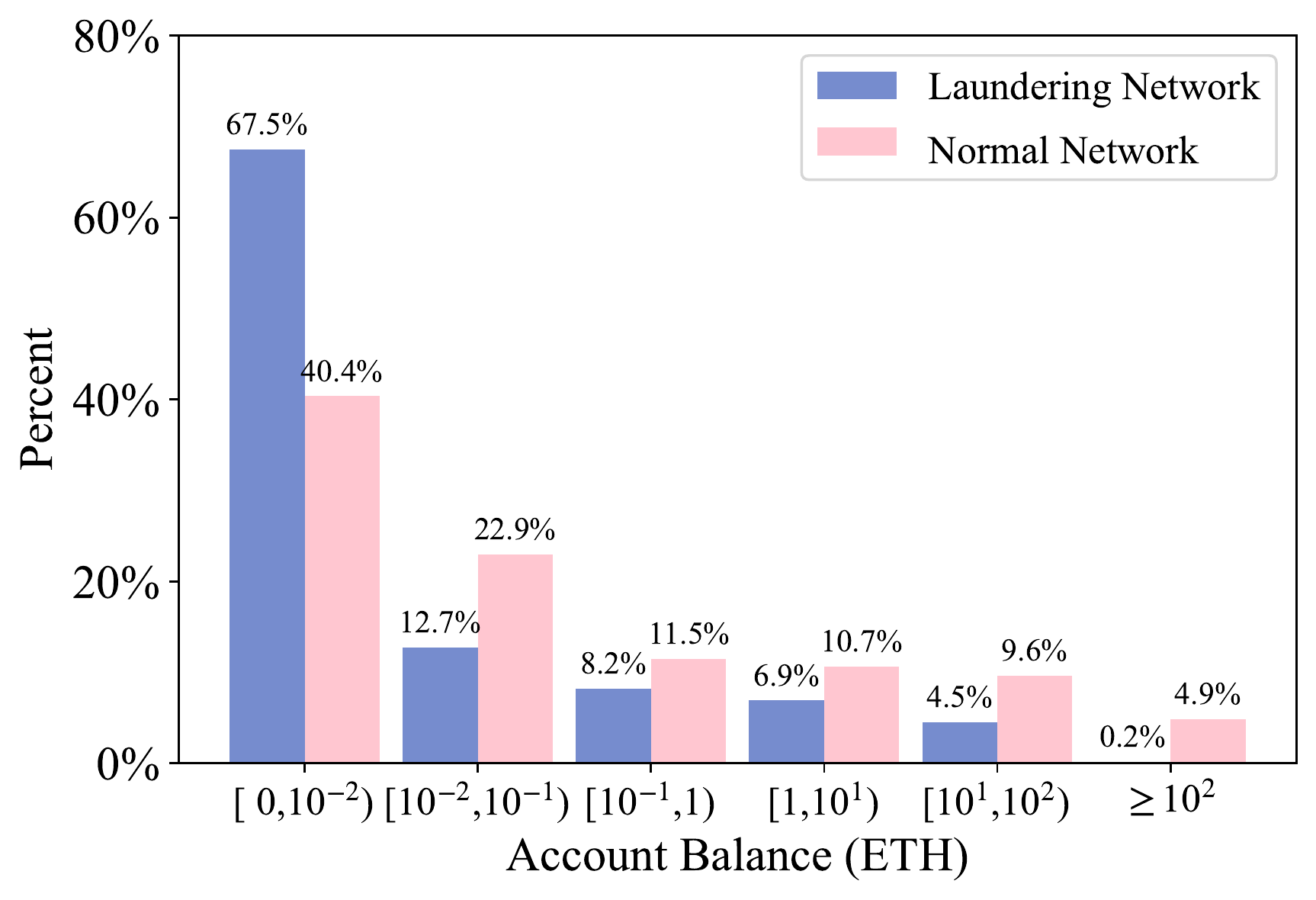}
  \caption{Account balance on normal network and money laundering network. }
  \label{fig:Ac balance}
\end{figure}
%---------------介绍账户余额指标公式-------------
%为了和Trait 3对比去探究以太坊洗钱账户是否是快进快出账户，我们引入了平均交易间隔去measure洗钱账户的交易速度。
%账户的余额是+公式，

To compare with Trait 3 to find whether ML accounts on Ethereum are zero out middle accounts, we put forward the account balance, and the account balance of an account $u$ is defined as 
\begin{equation}
\nonumber
    B(u) = In(u)-Out(u),(u \in N),
\label{equ:balance}
\end{equation}
where $In(u)$ is the sum amount of ETH  transferred to $u$ and $Out(u)$ is the sum amount of ETH transferred out from $u$. 

%-----------------账户余额结果分析--------------
%(1)只看ML的余额分布,88.41%的UpBitHacker 的账户余额小于1，这说明洗钱账户不会把很多钱留在自己的钱包里，
%(2)与普通账户相比，账户余额在[0,0.01)的Upbit  Hacker的占比比Normal Account的大，但账户余额大于0.01的UpbitHacker的占比都比Normal Account的小
%这些都说明大多数洗钱账户不会将很多钱留在账户里，因为转出几乎所有的转入的钱不仅可以最大化他们提现的数额，还会减小被发现的风险。

We divide the account balance into 6 ranges, then calculate the percentage of each balance range of ML accounts and normal accounts. Fig. \ref{fig:Ac balance} shows the percentage of account balance of ML accounts and normal accounts.  We can find: (i) Comparing the percentage of different balance ranges of ML accounts, the balance of 88.4\% ML accounts is less than 1 ETH; (ii) Comparing with normal accounts, the percentage of ML accounts with an account balance of [0, $10^{-2}$) is larger than that of normal Account, while the percentage of ML accounts with an account balance greater than $10^{-2}$ is smaller than that of normal accounts. Those findings indicate most ML accounts do not prefer to keep lots of money because transferring almost all incoming money out not only maximizes the amount of their withdrawals but also reduces the risk of being discovered.

%举账户余额为0的账户例子
%0x0a29073f68fb0962A527996a37e85d4ed8504a50 
%典型的zero out middle 账户，24个交易对，每次收到一笔钱都会全部转出，导致余额为0.
\textbf{[Example]} In ML dataset, 0x0a290\footnote{\scriptsize\url{https://cn.etherscan.com/txs?a=0x0a29073f68fb0962a527996a37e85d4ed8504a50}} is a classic zero out middle account with 24 transaction pairs. Every time the account received a sum of money, it would all be transferred out, resulting in a balance of 0.

\begin{mdframed}
\textbf{Finding 3.} Like traditional ML accounts,  ML accounts on Ethereum are zero out middle accounts. They transfer almost all incoming money out to potentially benefit in a big way.
\end{mdframed}

% \subsection{How dense is $\small\textsf{LaunderNet}$, $\small\textsf{LaunderEtherNet}$, and $\small\textsf{LaunderTokenNet}$?}
\subsection{How dense are ML networks on Ethereum?}
%---------------介绍网络密度指标公式----------------------
%为了和Trait 4对比以探究以太坊洗钱网络是否也很密集，我们引入了密度指标。网络密度的计算公式为+公式
%qs-comment: 公式加引用
To compare with Trait 4 and see if the ML network on Ethereum is also dense, we use density as an indicator. The density of a network $G$ is defined as~\cite{Lee2020Measurements}
\begin{equation}
\nonumber
    D(G) = \frac{4|E|}{|N|(|N|-1)} ,
\label{equ:density}
\end{equation}
where $|N|$ is the number of nodes and $|E|$ is the number of edges in network G.

%---------------网络密度分析-----------------
%计算几个网络的密度并进行比较后，见表X，我们发现：
%(1)MLETH _Net的密度远大于正常的$\textsf{\small{TransactionNet}}$;并且MLToken_Net的密度远大于正常的$\textsf{\small{TransactionNet}}$。 这足以说明区块链洗钱网络是密集的，这是因为参与区块链洗钱的账户进行了频繁的转入转出交易。
%(2)MLETH_Net网络密度略小于MLToken _Net网络的密度。 这是因为非法分子试图通过将ETH转换成多种类型的代币进行洗钱，例如XX交易，就是将多少个ETH转换成了XX种代币。这就会导致MLToken密度大于ETH的。

%qs-comment 这段的段头改了
%qs-comment 缺将ETH转换为其他代币的例子，我一定要找到！
The densities of several networks are shown in Table \ref{tab:density}, the $\textsf{\small{$\textsf{\small{TransactionNet}}$}}$ and $\textsf{\small{TokenNet}}$ are the normal network trading in ETH and in tokens respectively. We observe from Table \ref{tab:density}: (i) The density of $\textsf{\small{LaunderEtherNet}}$ is much higher than that of $\textsf{\small{TransactionNet}}$, and the density of $\textsf{\small{LaunderTokenNet}}$ is much higher than that of $\textsf{\small{TokenNet}}$. This is enough to show that the ML network on Ethereum is dense, because the ML accounts carry out frequent transfer transactions. (ii) The $\textsf{\small{LaunderEtherNet}}$ density is slightly less than the $\textsf{\small{LaunderTokenNet}}$ density. This is because illegal 
accounts complete multiple token transactions by calling a smart contract, which makes the  $\textsf{\small{LaunderTokenNet}}$  denser.  For example, 0x41f5f\footnote{\scriptsize{0x41f5f63f1c4d5cd0d58d7dc19efcb1f6b5f779ca59a82a593697ac80d56eeb90}} is a transaction of calling a smart contract which leads to four token transacions.

%Finding4: 和传统的洗钱网络类似，区块链洗钱网络也很密集。并且狡猾的犯罪分子还会将ETH转换为多种代币来达到混淆的目的。
\begin{mdframed}
\textbf{Finding 4.} Like traditional ML networks, ML networks on Ethereum are dense. At the same time, cunning criminals will complete multiple token transactions by calling a smart contract.
\end{mdframed}

% % Table generated by Excel2LaTeX from sheet 'Sheet1'
% \begin{table*}[htbp]
% \footnotesize
%   \centering
%   \caption{Density of networks}
%     \begin{tabular}{lrrrrr}
%     \toprule
%         \multicolumn{1}{l}{Networks}
%           & \multicolumn{1}{l}{ML\_Net} & \multicolumn{1}{l}{MLETH\_Net} & \multicolumn{1}{l}{$\textsf{\small{TransactionNet}}$} & \multicolumn{1}{l}{MLToken\_Net} & \multicolumn{1}{l}{TokenNet} \\
%     \midrule
%     Density & 1.41E-05 & 2.28E-05 & 1.87E-07 & 2.51E-05 & 1.87E-07 \\
%     \bottomrule
%     \end{tabular}%
%   \label{tab:density}%
% \end{table*}%


% \begin{table}[htbp]
% \footnotesize
% \tabcolsep=0.03cm
%   \centering
%   \caption{Density of networks}
%   \begin{tabular}{lccccc}
%         \toprule
%     \textbf{Networks} & \multicolumn{1}{p{4.19em}}{{Launder}\newline{}{Net}}  & \multicolumn{1}{p{5.5em}}{{LaunderEther}\newline{}{Net}} & \multicolumn{1}{p{5.8em}}{{TansactionNet}\newline{}{Net~\cite{Lee2020Measurements}}}& \multicolumn{1}{p{5.8em}}{{LaunderToken}\newline{}{Net}} & \multicolumn{1}{p{4em}}{{TokenNet}\newline{}{\cite{Lee2020Measurements}}} \\
%     \midrule
%     \textbf{Density}  & 1.41E-05 & 2.28E-05 & 1.87E-07 & 2.51E-05 & 1.87E-07 \\
%         \bottomrule
%     \end{tabular}%
%   \label{tab:density}%
% \end{table}%

\begin{table}[htbp]
\footnotesize
\tabcolsep=0.03cm
  \centering
  \caption{Density of networks}
  \begin{tabular}{lccccc}
        \toprule
    \textbf{Networks} & \multicolumn{1}{p{4.19em}}{\makecell[c]{$\textsf{\scriptsize{Launder}}$}\newline{}\makecell[c]{$\textsf{\scriptsize{Net}}$}}  & \multicolumn{1}{p{5.5em}}{\makecell[c]{$\textsf{\scriptsize{LaunderEther}}$}\newline{}\makecell[c]{$\textsf{\scriptsize{Net}}$}} & \multicolumn{1}{p{5.8em}}{\makecell[c]{$\textsf{\scriptsize{Tansaction}}$}\newline{}\makecell[c]{$\textsf{\scriptsize{Net~\cite{Lee2020Measurements}}}$}} & \multicolumn{1}{p{5.8em}}{\makecell[c]{$\textsf{\scriptsize{LaunderToken}}$}\newline{}\makecell[c]{$\textsf{\scriptsize{Net}}$}} & \multicolumn{1}{p{4em}}{\makecell[c]{$\textsf{\scriptsize{TokenNet}}$}\newline{}\makecell[c]{\cite{Lee2020Measurements}}} \\
    \midrule
    \textbf{Density}  & 1.41E-05 & 2.28E-05 & 1.87E-07 & 2.51E-05 & 1.87E-07 \\
        \bottomrule
    \end{tabular}%
  \label{tab:density}%
\end{table}%
\subsection{How different is  closeness centrality on ML subsets?}

%-----------------介绍中心性公式和定义------------------
%为了和Trait 5对,我们使用closeness centrality探究了区块链洗钱的2个子网的特征。closeness centrality 是connected网络中each节点到其他节点的average shortest path distance的倒数，定义为...
%qs-comment 公式加引用
In order to compare with Trait 5, we explore the characteristics of integration subset of ML network by closeness centrality. We estimate the level of closeness centrality of integration subset through comparing it with that of layering subset. Closeness centrality\footnote{\scriptsize{\url{https://networkx.org/documentation/stable/reference}}} is the reciprocal of the average shortest path distance of an account u to all others. Formally,
\begin{equation}
\nonumber
    C(u) = \frac{|N|-1}{\sum_{v = 1}^{|N|-1} dist(u,v)}, (u,v \in N),
\label{equ:density}
\end{equation}

The closeness centrality statistics of the two subsets measured are shown in table \ref{tab:closeness}. By analyzing these data, we find: The average of closeness centrality in the integration subset is lower than that in the layering subset. The reason is that there are various types of service providers on Ethereum, for example, gambling, lending platform, centralized exchange, decentralized exchange, and coin mixing service, etc., which provide criminals with many options to withdraw cash. In order to reduce the risk of exposure, accounts will be extra cautious to convert cash from multiple service providers in the integration subset, and the average shortest path distance of each account to others is longer, resulting in closeness centrality not high. But in traditional ML, they can only aggregate money into several accounts because of the limited number of withdrawal platforms and accounts available, and the average shortest path distance of each account to others is shorter, leading to high closeness centrality.

%Finding 5：和传统的洗钱网络不同，由于区块链上的去中心化驱动了motivate各式各样的服务商，所以区块链洗钱聚合子网的closeness 均值并没有远大于layering子网。 另一方面，犯罪分子仍然会更倾向于使用一些著名的服务商进行提现。
\begin{mdframed}
\textbf{Finding 5.} Different from traditional ML networks, the closeness centrality of integration subset is not much high, because the decentralization flourishes various service providers on Ethereum.
\end{mdframed}

% \begin{table}[htbp]
% \tabcolsep=0.08cm
% \small
%   \centering
%   \caption{Closeness Centrality}
%     \begin{tabular}{lp{13.565em}ll}
%     \toprule
%     subset & \multicolumn{1}{l}{Description} & Avg. C & Range of C \\
%     \midrule
%     Layering & \multicolumn{1}{{p{12em}}}{Disperse ill-gotten money\newline{} from source accounts}   & 2.69E-02 & 4.86E-02 \\
%     Integration & \multicolumn{1}{{p{12em}}}{Integrate dispersive money \newline{}into service providers} & 2.66E-02 & 5.45E-02 \\
%     \bottomrule
%     \end{tabular}%
%   \label{tab:closeness}%
% \end{table}%

\begin{table}[htbp]
\tabcolsep=0.08cm
\footnotesize
  \centering
  \caption{Avg. closeness centrality of ML subsets.}
    \begin{tabular}{lp{13.565em}ll}
    \toprule
    \textbf{Subset} & \multicolumn{1}{l}{\textbf{Description}}  & \textbf{Avg. closeness centrality} \\
    \midrule
    Layering & \multicolumn{1}{{p{12em}}}{Accounts dispersing \newline{}ill-gotten money from source accounts}   & 5.01E-06 \\
    Integration & \multicolumn{1}{{p{12em}}}{Accounts integrating \newline{}dispersive money into service providers}  & \textbf{1.46E-06} \\
    \bottomrule
    \end{tabular}%
  \label{tab:closeness}%
\end{table}%

%\multicolumn{1}{{p{4em}}}{\textbf{Duration\newline{} (Day)}}

\section{Conclusions and Future Work}
\label{sec:conclusion}
%结论
%从网络性质的角度系统的研究洗钱网络是否有传统的洗钱特性。
%收集Upbit案件的洗钱交易作为以太坊上的洗钱数据集
%基于数据集，我们收获了有趣的发现，通过分析了5个网络性质
%以太坊上的洗钱网络和传统洗钱网络有相似的特性，如XX，但也存在不同的网络特性，如XX
%这些发现能帮助更好了解区块链上的洗钱。
In this paper, we conducted a systematic study to analyze whether the ML networks on Ethereum have traditional traits of ML from the perspectives of network properties. We collect ML transaction of \textit{Upbit Hack} event as an example of the ML dataset on Ethereum. Based on the dataset, we acquire a series of interesting findings by analyzing five network properties. The ML networks on Ethereum have some similar traits with the traditional ML networks, such as fast velocity of transfer, zero balance of accounts, and high density of network. While they have several different traits with the traditional ML networks, for example, the ML accounts on Ethereum prefer to transfer small-volume money and the closeness centrality of integration subset is not high. These findings contributes to a better understanding of the ML on Ethereum.
  
%未来工作
%1.根据以太坊上洗钱网络的特性去用在下游任务的提取特征中,这会使得洗钱交易/账户的识别具有更高准确率。
%2.有潜力去提出一个以太坊上的反洗钱检测框架，参考传统的AML框架，并且以太坊洗钱账户的特性可以被用来设计检测框架的异常指标。
%3.时序分析洗钱网络的演化，更能了解以太坊洗钱网络上从开始洗钱到洗钱完成的整个过程
%4.相似的分析可以被用在代币交易网络中去揭露有趣的现象和特性。
\textbf{Future work.} Following our insights into ML networks on Ethereum, there are ample opportunities for future work: (i) By characterizing the money laundering (ML) networks on Ethereum, the analysis can be used for feature extraction in downstream tasks such as account classification and son on~\cite{huang2022account,Liu2022FAGNN,mixing}, which may lead to more accurate detection of ML transactions and accounts; (ii) It is potential to propose a AML detection frame on Ethereum referring to the traditional AML detection frame of existing work~\cite{li2020flowscope,sun2021cubeflow}, and the traits of ML accounts on Ethereum can be leveraged to design abnormal indicators of the detection frame; (iii) Further analysis of the evolution of ML networks on Ethereum from a temporal perspective would be interesting~\cite{lu2022evolution,huang2022temporal}, as would have better understanding of the entire process of ML on Ethereum, from the start of the laundering process to its completion; (iv) A similar line of measurements and analyses can be applied to token transaction networks~\cite{chen2020traveling} to unearth interesting phenomena and traits on ML networks.
%To facilitate such research directions, we open source our datasets: https://github.com/XXX. Quite naturally 

\section{Acknowledgment}
The work described in this paper is supported by the National Natural Science Foundation of China (61973325), the Natural Science Foundations of Guangdong Province (2021A1515011661), and the Guangzhou Basic and Applied Basic Research Project (202102020616).

\bibliographystyle{IEEEtran}
\bibliography{conference_101719}

\begin{thebibliography}{10}
\providecommand{\url}[1]{#1}
\csname url@samestyle\endcsname
\providecommand{\newblock}{\relax}
\providecommand{\bibinfo}[2]{#2}
\providecommand{\BIBentrySTDinterwordspacing}{\spaceskip=0pt\relax}
\providecommand{\BIBentryALTinterwordstretchfactor}{4}
\providecommand{\BIBentryALTinterwordspacing}{\spaceskip=\fontdimen2\font plus
\BIBentryALTinterwordstretchfactor\fontdimen3\font minus
  \fontdimen4\font\relax}
\providecommand{\BIBforeignlanguage}[2]{{%
\expandafter\ifx\csname l@#1\endcsname\relax
\typeout{** WARNING: IEEEtran.bst: No hyphenation pattern has been}%
\typeout{** loaded for the language `#1'. Using the pattern for}%
\typeout{** the default language instead.}%
\else
\language=\csname l@#1\endcsname
\fi
#2}}
\providecommand{\BIBdecl}{\relax}
\BIBdecl

\bibitem{nakamoto2008bitcoin}
S.~Nakamoto, ``Bitcoin: A peer-to-peer electronic cash system bitcoin: A
  peer-to-peer electronic cash system,''
  \emph{https://bitcoin.org/bitcoin.pdf}, 2008.

\bibitem{lin2022ethereum}
D.~Lin, J.~Wu, Q.~Xuan, and K.~T. Chi, ``Ethereum transaction tracking:
  {I}nferring evolution of transaction networks via link prediction,''
  \emph{Physica A: Statistical Mechanics and its Applications}, vol. 600, p.
  127504, 2022.

\bibitem{Lin2021Evolution}
D.~Lin, J.~Chen, J.~Wu, and Z.~Zheng, ``Evolution of ethereum transaction
  relationships: {T}oward understanding global driving factors from microscopic
  patterns,'' \emph{IEEE Transactions on Computational Social Systems}, vol.~9,
  no.~2, pp. 559--570, 2022.

\bibitem{Wu2021JNCA}
J.~Wu, J.~Liu, Y.~Zhao, and Z.~Zheng, ``{Analysis of cryptocurrency
  transactions from a network perspective: An overview},'' \emph{Journal of
  Network and Computer Applications}, vol. 190, p. 103139, 2021.

\bibitem{Chainalysis2022}
\BIBentryALTinterwordspacing
{Chainalysis Team}. (2022) The chainalysis 2022 crypto crime report. [Online].
  Available: \url{https://go.chainalysis.com/2022-crypto-crime-report.html}
\BIBentrySTDinterwordspacing

\bibitem{slowmist2022}
\BIBentryALTinterwordspacing
{SlowMist Team}. (2022) First half of the 2022 blockchain security and
  anti-money laundering analysis report. [Online]. Available:
  \url{https://www.slowmist.com/report/first-half-of-the-2022-report.pdf}
\BIBentrySTDinterwordspacing

\bibitem{CipherTrace2022}
\BIBentryALTinterwordspacing
{CipherTrace Team}. (2022) Cryptocurrency crime and anti-money laundering
  report. [Online]. Available:
  \url{https://ciphertrace.com/crime-and-anti-money-laundering-report-october-2022/}
\BIBentrySTDinterwordspacing

\bibitem{fatf2022}
\BIBentryALTinterwordspacing
FATF. (2022) Money laundering and terrorist financing red flag indicators
  associated with virtual assets. [Online]. Available:
  \url{http://www.fatf-gafi.org/publications/fatfrecommendations/documents/Virtual-Assets-Red-Flag-Indicators.html}
\BIBentrySTDinterwordspacing

\bibitem{sun2021cubeflow}
X.~Sun, J.~Zhang, Q.~Zhao, S.~Liu, J.~Chen, R.~Zhuang, H.~Shen, and X.~Cheng,
  ``Cubeflow: Money laundering detection with coupled tensors,'' in
  \emph{Proceedings of the Pacific-Asia Conference on Knowledge Discovery and
  Data Mining}, 2021, pp. 78--90.

\bibitem{li2020flowscope}
X.~Li, S.~Liu, Z.~Li, X.~Han, C.~Shi, B.~Hooi, H.~Huang, and X.~Cheng,
  ``Flowscope: Spotting money laundering based on graphs,'' in
  \emph{Proceedings of the AAAI conference on artificial intelligence},
  vol.~34, no.~04, 2020, pp. 4731--4738.

\bibitem{sun2022monlad}
X.~Sun, W.~Feng, S.~Liu, Y.~Xie, S.~Bhatia, B.~Hooi, W.~Wang, and X.~Cheng,
  ``Monlad: Money laundering agents detection in transaction streams,'' in
  \emph{Proceedings of the ACM International Conference on Web Search and Data
  Mining}, 2022, pp. 976--986.

\bibitem{mahootiha2021designing}
M.~Mahootiha, A.~H. Golpayegani, and B.~Sadeghian, ``Designing a new method for
  detecting money laundering based on social network analysis,'' in
  \emph{Proceedings of the International Computer Conference, Computer Society
  of Iran}, 2021, pp. 1--7.

\bibitem{lokanan2022predicting}
M.~E. Lokanan, ``Predicting money laundering using machine learning and
  artificial neural networks algorithms in banks,'' \emph{Journal of Applied
  Security Research}, pp. 1--25, 2022, to be published,
  doi:\url{10.21203/rs.3.rs-2530874/v1}.

\bibitem{singh2019anti}
K.~Singh and P.~Best, ``Anti-money laundering: using data visualization to
  identify suspicious activity,'' \emph{International Journal of Accounting
  Information Systems}, vol.~34, p. 100418, 2019.

\bibitem{jullum2020detecting}
M.~Jullum, A.~L{\o}land, R.~B. Huseby, G.~{\AA}nonsen, and J.~Lorentzen,
  ``Detecting money laundering transactions with machine learning,''
  \emph{Journal of Money Laundering Control}, vol.~23, no.~1, pp. 173--186,
  2020.

\bibitem{wu2022transaction}
Z.~Wu, J.~Liu, J.~Wu, and Z.~Zheng, ``Transaction tracking on blockchain
  trading systems using personalized pagerank,'' \emph{doi preprint
  doi:2201.05757}, 2022.

\bibitem{Lee2020Measurements}
X.~T. Lee, A.~Khan, S.~{Sen Gupta}, Y.~H. Ong, and X.~Liu, ``{Measurements,
  analyses, and insights on the entire Ethereum blockchain network},'' in
  \emph{Proceedings of the Web conferenec}, 2020, pp. 155--166.

\bibitem{trans2vec2020Wu}
J.~Wu, Q.~Yuan, D.~Lin, W.~You, W.~Chen, C.~Chen, and Z.~Zheng, ``Who are the
  phishers? {P}hishing scam detection on ethereum via network embedding,''
  \emph{IEEE Transactions on Systems, Man, and Cybernetics: Systems}, vol.~52,
  no.~2, pp. 1156--1166, 2022.

\bibitem{huang2022account}
T.~Huang, D.~Lin, and J.~Wu, ``Ethereum account classification based on graph
  convolutional network,'' \emph{IEEE Transactions on Circuits and Systems II:
  Express Briefs}, vol.~69, no.~5, pp. 2528--2532, 2022.

\bibitem{Liu2022FAGNN}
J.~Liu, J.~Zheng, J.~Wu, and Z.~Zheng, ``Fa-gnn: Filter and augment graph
  neural networks for account classification in ethereum,'' \emph{IEEE
  Transactions on Network Science and Engineering}, vol.~9, no.~4, pp.
  2579--2588, 2022.

\bibitem{lu2022evolution}
D.~Lu, J.~Liu, S.~Zhao, and J.~Wu, ``Evolution of locality on ethereum
  transaction network,'' in \emph{Proceedings of the IEEE International
  Symposium on Circuits and Systems}, 2022, pp. 3527--3531.

\bibitem{huang2022temporal}
B.~Huang, J.~Liu, J.~Wu, Q.~Li, and H.~Lin, ``Temporal analysis of transaction
  ego networks with different labels on ethereum,'' in \emph{Proceedings of the
  IEEE International Symposium on Circuits and Systems}, 2022, pp. 3517--3521.

\bibitem{chen2020traveling}
W.~Chen, T.~Zhang, Z.~Chen, Z.~Zheng, and Y.~Lu, ``Traveling the token world: A
  graph analysis of ethereum erc20 token ecosystem,'' in \emph{Proceedings of
  The Web Conference}, 2020, pp. 1411--1421.

\bibitem{modeling2020lin}
D.~Lin, J.~Wu, Q.~Yuan, and Z.~Zheng, ``Modeling and understanding ethereum
  transaction records via a complex network approach,'' \emph{IEEE Transactions
  on Circuits and Systems II: Express Briefs}, vol.~67, no.~11, pp. 2737--2741,
  2020.

\bibitem{mixing}
J.~Wu, J.~Liu, W.~Chen, H.~Huang, Z.~Zheng, and Y.~Zhang, ``Detecting mixing
  services via mining bitcoin transaction network with hybrid motifs,''
  \emph{IEEE Transactions on Systems, Man, and Cybernetics: Systems}, vol.~52,
  no.~4, pp. 2237--2249, 2022.

\end{thebibliography}

\end{document}